# Dynamical Casimir-Polder Force between a Metallic Particle and a Perfectly Conducting Plate

G.V. Dedkov[1] and A.A. Kyasov

Nanoscale Physics Group, Kabardino-Balkarian State University, Nalchik, Russian Federation

For the first time, we have studied the impact of magnetic polarizability of a neutral metallic particle on the dynamical Casimir-Polder force, when the particle is uniformly moving with relativistic velocity parallel to the surface of an ideally conducting plate. The material properties of particle are taken into account within the Drude model approach.

## 1. Introduction

Some years ago, the impact of magnetic polarizability of a small conducting electrically neutral particle on the attractive Casimir-Polder force has been studied in our work [1], involving a static case. The fluctuating magnetic moment appears even on a resting nonmagnetic object due to the random Foucault currents which are induced by the non-stationary or random external magnetic fields penetrating the volume of the object.

As well, in the recent papers [2,3] we have considered the dynamical Casimir-Polder force between a moving neutral atom and a metallic surface (a perfectly conducting and permeable plate). The atom was assumed to be moving parallel to the surface with relativistic velocity $V$. We have found a very strong influence of the velocity (relativistic energy) on the Casimir-Polder force in the case of an electrically polarizable atom. In general, the increase of the velocity factor $\beta = V/c$ ($c$ is the speed of light in vacuum) or the energy factor $\gamma = (1-\beta^2)^{-1/2}$ results in a significant decrease of the Casimir-Polder force $F_z$, whereas the dependence $F_z(\beta)$ is nonmonotonous and has several interesting features.

It is quite natural to consider the dynamical Casimir-Polder force on the moving magnetically polarizable particle (a small metal ball). In this paper we restrict our consideration assuming the surface to be ideally conducting. The particle is assumed to be described by the Drude approximation for the dielectric permittivity.

---

[1] Corresponding author e-mail: gv_dedkov@mail.ru

## 2. General theoretical expressions

To start with we recall our general formula for the dynamical Casimir-Polder force between a magnetically and electrically polarizable particle and the surface of a thick plate (semiinfinite homogeneous medium with flat boundary), corresponding to the dipole approximation of fluctuation electrodynamics [3,4]

$$F_z = -\frac{\hbar \gamma}{2\pi^2} \int_0^\infty d\omega \int_{-\infty}^{+\infty} dk_x \int_{-\infty}^{+\infty} dk_y \left\{ \begin{array}{l} \alpha_m''(\gamma\omega^+)\mathrm{Re}[\exp(-2q_0 z_0) R_m(\omega,\mathbf{k})]\coth\left(\frac{\gamma\hbar\omega^+}{2k_B T_1}\right) + \\ + \alpha_m'(\gamma\omega^+)\mathrm{Im}[\exp(-2q_0 z_0) R_m(\omega,\mathbf{k})]\coth\left(\frac{\hbar\omega}{2k_B T_2}\right) + (m \to e) \end{array} \right\} \quad (1)$$

$$R_m(\omega,\mathbf{k}) = \Delta_m(\omega)\left[2(k^2 - k_x^2 \beta^2)(1 - \omega^2/k^2 c^2) + (\omega^+)^2/c^2\right] + \\ + \Delta_e(\omega)\left[2k_y^2 \beta^2 (1 - \omega^2/k^2 c^2) + (\omega^+)^2/c^2\right] \quad (2)$$

$$R_e(\omega,\mathbf{k}) = \Delta_e(\omega)\left[2(k^2 - k_x^2 \beta^2)(1 - \omega^2/k^2 c^2) + (\omega^+)^2/c^2\right] + \\ + \Delta_m(\omega)\left[2k_y^2 \beta^2 (1 - \omega^2/k^2 c^2) + (\omega^+)^2/c^2\right] \quad (3)$$

$$\Delta_e(\omega) = \frac{q_0 \varepsilon(\omega) - q}{q_0 \varepsilon(\omega) + q}, \quad \Delta_m(\omega) = \frac{q_0 \mu(\omega) - q}{q_0 \mu(\omega) + q}, \quad q = \left(k^2 - (\omega^2/c^2)\varepsilon(\omega)\mu(\omega)\right)^{1/2}, \\ q_0 = (k^2 - \omega^2/c^2)^{1/2}, \quad k^2 = k_x^2 + k_y^2, \quad \omega^+ = \omega + k_x V \quad (4)$$

where $T_1$ and $T_2$ are the particle and surface temperatures, $z_0$ is the particle-surface separation, $\alpha_m(\omega)$ is magnetic polarizability of the particle, one-primed and double-primed quantities denote its real and imaginary parts, $\varepsilon(\omega)$ and $\mu(\omega)$ denote the bulk dielectric/magnetic permeabilities of the plate. The term $(e \to m)$ in the figure brackets of (3), corresponding to the contribution from electric polarizability of the particle, is obtained by replacing the incoming variables in the first two terms: $\alpha_m \to \alpha_e, R_m \to R_e, \Delta_m \leftrightarrow \Delta_e$. The global system of magnetodielectric bodies is assumed to be out of thermal equilibrium but in the stationary regime, the plate and surrounding vacuum background are at equilibrium with temperature $T_2$.

To examine the case of an ideally conducting plate, one must put $\Delta_e(\omega) = 1, \Delta_m(\omega) = -1$ (that involves $\varepsilon(\omega) \to \infty, \mu(\omega) = 1$). As well, we pass to the limit $T_1 \to 0, T_2 \to 0$, just as we have done in [2,3]. Then Eq. (1) reduces to the simpler form

$$F_z = F_m^{(0)} + F_m^{(1)} + F_e^{(0)} + F_e^{(1)} \quad (5)$$

$$F_m^{(0)} = \frac{\hbar}{\pi^2 \gamma} \int_{-\infty}^{+\infty} dk_x \int_{-\infty}^{+\infty} dk_y \int_0^{\infty} d\xi \exp(-2\sqrt{k^2 + \xi^2/c^2}\, z_0) \cdot \mathrm{Im}[i \cdot \alpha_m(\gamma(i\xi + k_x V))] \cdot (k^2 + \xi^2/c^2) \quad (6)$$

$$F_m^{(1)} = -\frac{4\hbar}{\pi^2 \gamma} \int_0^{\infty} dk_x \int_0^{\infty} dk_y \int_0^{k_x V} d\omega \exp(-2\sqrt{k^2 - \omega^2/c^2}\, z_0) \mathrm{Im}[\alpha_m(\gamma(\omega - k_x V))] \cdot (k^2 - \omega^2/c^2) \quad (7)$$

$$F_e^{(0)} = -\frac{\hbar}{\pi^2 \gamma} \int_{-\infty}^{+\infty} dk_x \int_{-\infty}^{+\infty} dk_y \int_0^{\infty} d\xi \exp(-2\sqrt{k^2 + \xi^2/c^2}\, z_0) \cdot \mathrm{Im}[i \cdot \alpha_e(\gamma(i\xi + k_x V))] \cdot (k^2 + \xi^2/c^2) \quad (8)$$

$$F_e^{(1)} = \frac{4\hbar}{\pi^2 \gamma} \int_0^{\infty} dk_x \int_0^{\infty} dk_y \int_0^{k_x V} d\omega \exp(-2\sqrt{k^2 - \omega^2/c^2}\, z_0) \mathrm{Im}[\alpha_e(\gamma(\omega - k_x V))] \cdot (k^2 - \omega^2/c^2) \quad (9)$$

We see that the only difference between Eqs.(6),(7) describing the contributions of magnetic polarizability of the particle and Eqs.(8), (9) describing the electric contributions, is the opposite sign.

Prior to consider the general case of a permeable particle, it is worthwhile to examine the simplest case of an ideally conducting particle. In this case we must put $\alpha_m(\omega) = -R^3/2$ and $\alpha_e(\omega) = R^3$ in Eqs.(6)--(9) [5], where $R$ is the particle radius. Integrating Eqs. (6) and (8) yields

$$F_m^{(0)} = -\frac{3\hbar c R^3}{4\pi z_0^5 \gamma}, \quad F_e^{(0)} = -\frac{3\hbar c R^3}{2\pi z_0^5 \gamma} \quad (10)$$

In this case $F_m^{(1)} = 0, F_e^{(1)} = 0$ since $\alpha_{m,e}(\omega)$ are real, and the resulting Casimir-Polder force is given by

$$F_z = -\frac{9\hbar c R^3}{4\pi z_0^5 \gamma} \quad (11)$$

In the nonrelativistic case at $\gamma \approx 1, \beta \ll 1$

$$F_z = -\frac{9\hbar c R^3}{4\pi z_0^5}\left(1 - \frac{\beta^2}{2}\right), \quad (12)$$

In the static case $\gamma = 1, \beta = 0$, correspondingly,

$$F_z = -\frac{9\hbar c R^3}{4\pi z_0^5} \quad (13)$$

Eq.(13) is in complete agreement with a quantum electrodynamic calculation [7]. It is worth noting that the used dipole approximation implies $R \ll z_0$.

According to Eqs. (11), (12) the dynamical Casimir-Polder force slightly decreases (by modulus) with increasing velocity factor at $\beta < 1$ and becomes inversely proportional to the relativistic factor $\gamma$ at $\beta \to 1$. The sign of force corresponds to attraction in the whole range of separations.

## 3. A permeable metallic particle and an ideal plate

Now we pass to the case where the particle is permeable and its dielectric permittivity is described by the Drude-like function

$$\varepsilon(\omega) = 1 - \frac{\omega_p^2}{\omega(\omega + i/\tau)} \tag{14}$$

In what follows we will use parameters $\omega_p = 1.37 \cdot 10^{16}$ $rad/s$, $\tau = 1.89 \cdot 10^{-14}$ $s$ corresponding to gold. According to [6], the magnetic and electric polarizabilities of conducting ball ($\sigma = \omega_p^2 \tau / 4\pi$) are given by

$$\alpha_m(\omega) = -\frac{R^3}{2}\left(1 - \frac{3}{x^2} + \frac{3}{x}\cot x\right) \equiv -\frac{R^3}{2}\chi_m(x),\ x = \frac{(1+i)}{c}\sqrt{2\pi\sigma\omega} \tag{15}$$

$$\alpha_e(\omega) = R^3 \frac{\varepsilon(\omega) - 1}{\varepsilon(\omega) + 2} = R^3 \frac{\omega_p^2}{\omega_p^2 - 3\omega^2 - 3\omega i/\tau} \tag{16}$$

Introducing new variables $k_x = x/2z_0, k_y = y/2z_0, \xi, \omega = \omega_p \Omega$, and denoting

$\lambda = 2\omega_p z_0 / c$, $\kappa = \omega_p \tau$, Eqs.(6)-(9) are rewritten in the form

$$F_m^{(0)} = -\frac{\hbar \omega_p R^3}{16\pi^2 z_0^4 \gamma} \int_0^\infty dx \int_0^\infty dy \int_0^\infty d\Omega \exp\left(-\sqrt{x^2 + y^2 + \lambda^2 \Omega^2}\right)(x^2 + y^2 + \lambda^2 \Omega^2) \cdot$$
$$\cdot \mathrm{Re}[\chi_m(x_1(x,\Omega)) + \chi_m(x_2(x,\Omega))] \tag{17}$$

$$x_1(x,\Omega) = \frac{\omega_p R}{c}(1+i)\sqrt{0.5\kappa\gamma(i\cdot\Omega + \beta x/\lambda)} \tag{18}$$

$$x_2(x,\Omega) = \frac{\omega_p R}{c}(1+i)\sqrt{0.5\kappa\gamma(i\cdot\Omega - \beta x/\lambda)} \tag{19}$$

$$F_m^{(1)} = \frac{\hbar \omega_p R^3}{8\pi^2 z_0^4 \gamma} \int_0^\infty dx \int_0^\infty dy \int_0^{\beta x/\lambda} d\Omega \exp\left(-\sqrt{x^2 + y^2 - \lambda^2 \Omega^2}\right)(x^2 + y^2 - \lambda^2 \Omega^2) \cdot \mathrm{Im}[\chi_m(x_3(x,\Omega))] \tag{20}$$

$$x_3(x,\Omega) = \frac{\omega_p R}{c}(1+i)\sqrt{0.5\kappa\gamma(i\cdot\Omega - \beta x/\lambda)} \tag{21}$$

$$F_e^{(0)} = -\frac{\hbar \omega_p R^3}{8\pi^2 z_0^4 \gamma} \int_0^\infty dx \int_0^\infty dy \int_0^\infty d\Omega \exp\left(-\sqrt{x^2 + y^2 + \lambda^2 \Omega^2}\right)(x^2 + y^2 + \lambda^2 \Omega^2)$$
$$\cdot \mathrm{Re}[\chi_e(x_4(x,\Omega)) + \chi_e(x_5(x,\Omega))] \tag{22}$$

$$F_e^{(1)} = \frac{\hbar \omega_p R^3}{4\pi^2 z_0^4 \gamma} \int_0^\infty dx \int_0^\infty dy \int_0^{\beta x/\lambda} d\Omega \exp\left(-\sqrt{x^2 + y^2 - \lambda^2 \Omega^2}\right)(x^2 + y^2 - \lambda^2 \Omega^2) \cdot \mathrm{Im}[\chi_e(x_6(x,\Omega))] \tag{23}$$

$$\chi_e(t) = \frac{1}{1 - 3t^2 - 3i \cdot t / \kappa} \tag{24}$$

$$x_4(x, \Omega) = \gamma(i \cdot \Omega + \beta x / \lambda) \tag{25}$$

$$x_5(x, \Omega) = \gamma(i \cdot \Omega - \beta x / \lambda) \tag{26}$$

$$x_6(x) = \gamma(\Omega - \beta x / \lambda) \tag{27}$$

Integrating Eqs.(17), (20), (22), (23) with respect to $y$ and substituting $u$ for $\Omega$ according to $x^2 \pm \Omega^2 \lambda^2 = u^2$, yields

$$\begin{aligned}F_m^{(0)} = &-\frac{\hbar c R^3}{32\pi^2 z_0^5 \gamma} \int_0^\infty dx \int_x^\infty du \frac{u^4}{\sqrt{u^2 - x^2}} \left(-\frac{d^3 K_0(u)}{du^3}\right) \cdot \\ &\cdot \mathrm{Re}\left[\chi_m\left(x_1(x, \frac{\sqrt{u^2 - x^2}}{\lambda})\right) + \chi_m\left(x_2(x, \frac{\sqrt{u^2 - x^2}}{\lambda})\right)\right]\end{aligned} \tag{28}$$

$$F_m^{(1)} = \frac{\hbar c R^3}{16\pi^2 z_0^5 \gamma} \int_0^\infty dx \int_{x/\gamma}^x du \frac{u^4}{\sqrt{x^2 - u^2}} \left(-\frac{d^3 K_0(u)}{du^3}\right) \mathrm{Im}\left[\chi_m\left(x_3(x, \frac{\sqrt{x^2 - u^2}}{\lambda})\right)\right] \tag{29}$$

$$\begin{aligned}F_e^{(0)} = &-\frac{\hbar c R^3}{16\pi^2 z_0^5 \gamma} \int_0^\infty dx \int_x^\infty du \frac{u^4}{\sqrt{u^2 - x^2}} \left(-\frac{d^3 K_0(u)}{du^3}\right) \cdot \\ &\cdot \mathrm{Re}\left[\chi_e\left(x_4(x, \frac{\sqrt{u^2 - x^2}}{\lambda})\right) + \chi_e\left(x_5(x, \frac{\sqrt{u^2 - x^2}}{\lambda})\right)\right]\end{aligned} \tag{30}$$

$$F_e^{(1)} = \frac{\hbar c R^3}{8\pi^2 z_0^5 \gamma} \int_0^\infty dx \int_{x/\gamma}^x du \frac{u^4}{\sqrt{x^2 - u^2}} \left(-\frac{d^3 K_0(u)}{du^3}\right) \cdot \mathrm{Im}\left[\chi_e\left(x_6(x, \frac{\sqrt{x^2 - u^2}}{\lambda})\right)\right] \tag{31}$$

where $K_0(u)$ is the Bessel function. The double integrals in (28)-(30) rapidly converge numerically, since the integrand functions have no singular points. Integral (21) is more tedious owing to the resonance character of $\mathrm{Im}\,\chi_e(x_6(x, \Omega))$. However, in the case of metal particle where $\kappa = \omega_p \tau \gg 1$ (particularly, for gold we have $\kappa = 259$), $F_e^{(0)}$ and $F_e^{(1)}$ prove to be practically independent of $\tau$. This allows one to use the limit $\tau \to \infty$ and simpler formulas for $F_e^{(0)}$ and $F_e^{(1)}$ which have been obtained in [3] involving the case of atom-surface interaction. The corresponding expressions (see Eqs. (26), (27) in Ref. 3) are then modified by replacing $\alpha(0) \to R^3$ and $\omega_0 \to \omega_p / \sqrt{3}$:

$$F_e^{(0)} = \frac{4\hbar \omega_p^5 R^3}{3^{5/3} \pi^2 c^4 \gamma^6} \int_0^\infty dx \int_0^\infty dy \frac{(1 - \beta^2 x^2 + y^2)(x^2 + y^2)^{3/2}}{[(1+\beta x)^2 + y^2][(1-\beta x)^2 + y^2]} \left[\frac{d^3 K_0(t)}{dt^3}\right]_{t = \tilde{\lambda}\sqrt{x^2 + y^2}} \tag{32}$$

$$F_e^{(1)} = \frac{\hbar \omega_p R^3}{8\pi\sqrt{3}\, z_0^{\,4} \gamma} \int_{\tilde{\lambda}/\beta}^{\infty} dx \frac{x^4}{\sqrt{x^2 + \tilde{\lambda}_0^{\,2}}} \frac{d^3 K_0(x)}{dx^3} \qquad (33)$$

where $\tilde{\lambda} = \lambda/\gamma\sqrt{3}$ and $\tilde{\lambda}_0 = \lambda/\sqrt{3}$. Using Eqs. (32), (33) makes it possible to calculate $F_e^{(0)}$ and $F_e^{(1)}$ at any values of $\beta$ and $\gamma$. Still simpler formulas in the limit cases $\beta \ll 1$ and $\gamma \gg 1$ are given in [3], as well.

## 4. Numerical results

Figures 1--4 display the results of our calculations according to Eqs.(28)—(31). All forces are normalized by the "etalon" force $F_0 = -9\hbar\, cR^3/4\pi z_0^{\,5}\gamma$ corresponding to Eq. (13), describing the Casimir-Polder between an ideally conducting ball and the surface. We use the notations

$$f_e = f_e^{(0)} + f_e^{(1)} = (F_e^{(0)} + F_e^{(1)})/F_0 \qquad (34)$$

$$f_m = f_m^{(0)} + f_m^{(1)} = (F_m^{(0)} + F_m^{(1)})/F_0 \qquad (35)$$

As it follows from calculations (Figs.1—3), the reduced components $f_e$, $f_m$ of the force turn out to be velocity-independent, despite that their separate components $f_{e,m}^{(0)}, f_{e,m}^{(1)}$ do. In addition, $f_e$ does not depend on the particle radius $R$ and thus it is solely the distance-dependent. The dotted line in Fig.1c shows a simple analytical fit to $f_e(z_0)$ of the form $f_e(z_0) \approx 0.67(1 - \exp(-0.0154 z_0))$, where $z_0$ is expressed in $nm$. Moreover, as we can see from Fig. 1, $f_e \to 0$ at $z_0 \to 0$. This is caused by the use of normalization factor $F_0$, because the electric part of the Casimir-Polder force in the case of Drude metal ball has the asymptotics $z_0^{-4}$ which is characteristic for the nonretarded Van der Waals force.

On the contrary, the magnetic contribution $f_m$ depends on both parameters $z_0$ and $R$. Some of these dependences are shown in Figs. 2-4 at different values of $z_0, R$. The negative sign of $f_m^{(1)}$ (Fig. 2a,b and Fig. 3a,b) means that this contribution to the Casimir-Polder force has the repulsive character. However, since $\left|f_m^{(1)}\right| < \left|f_m^{(0)}\right|$ the sum $f_m = f_m^{(0)} + f_m^{(1)}$ proves to be attractive, as $f_e$ does. Comparing the distance dependences for $f_e$ and $f_m$, we can note their principally different behavior: with increasing $z_0$ the former one tends to the asymptotic value 2/3 (one must divide Eq. (10) by $F_0$) while the latter monotonously goes down. The asymptotic value $f_m = 1/3$ is reached only at small particle-surface separations, where the dipole approximation is not valid. We can conclude that impact of material properties reduces (by modulus) the value of the total Casimir-Polder force as compared with an ideal case, Eq.(13). Fig.4 shows that the difference between the values of $f_m$ and the limiting result $f_m = 1/3$ decreases with increasing $R$.

## Conclusions

Generally, it turns out that the contribution of magnetic polarizability is smaller than the contribution of electric polarizability and the total dynamical Casimir-Polder force proves to be attractive in the whole range of particle-surface separations. Both electric and magnetic contributions to the force have a universal inverse dependence on the relativistic factor $\gamma$. With increasing the distance the electric contribution to the force monotonously tends to the asymptotic values in the case of an ideally conducting particle and a plate, whereas the magnetic contribution monotonously goes down. This results in a smaller final value of the Casimir-Polder (by modulus) as compared with an ideal case. The involved deviation becomes less significant with increasing particle radius.

FIGURE CAPTIONS

Fig.1. Dependences $f_e, f_e^{(0)}, f_e^{(1)}$ on $z_0$ at $\beta = 0.5$ (a) and $\beta = 0.99$ (b). The dashed, dashed-dotted and solid lines correspond to $f_e^{(0)}, f_e^{(1)}$ and $f_e$, the symbols correspond to static values of $f_e$ at $\beta = 0$. The solid line in case (c) corresponds to the calculated $f_e$, and the dashed line corresponds to the fitting function $f_e(z_0) \approx 0.67(1 - \exp(-0.0154 z_0))$.

Fig. 2 Dependences $f_m, f_m^{(0)}, f_m^{(1)}$ on the distance $z_0$ at $R = 1 nm$ and different $\beta$. The dashed dashed-dotted and solid lines in Figs. 2a,b correspond to $f_m^{(0)}, f_m^{(1)}$ and $f_m$.

Fig. 3. The same as in Fig. 2 at $R = 3$ nm.

Fig. 4. Dependence $f_m$ on $z_0$ and $R$.

# FIGURES

Fig. 1a

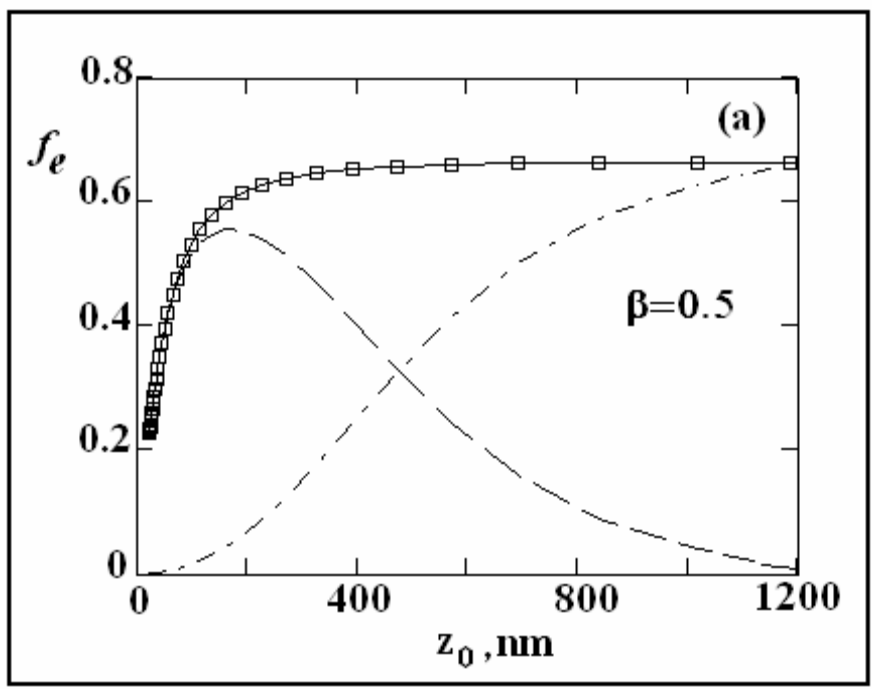

Fig.1b

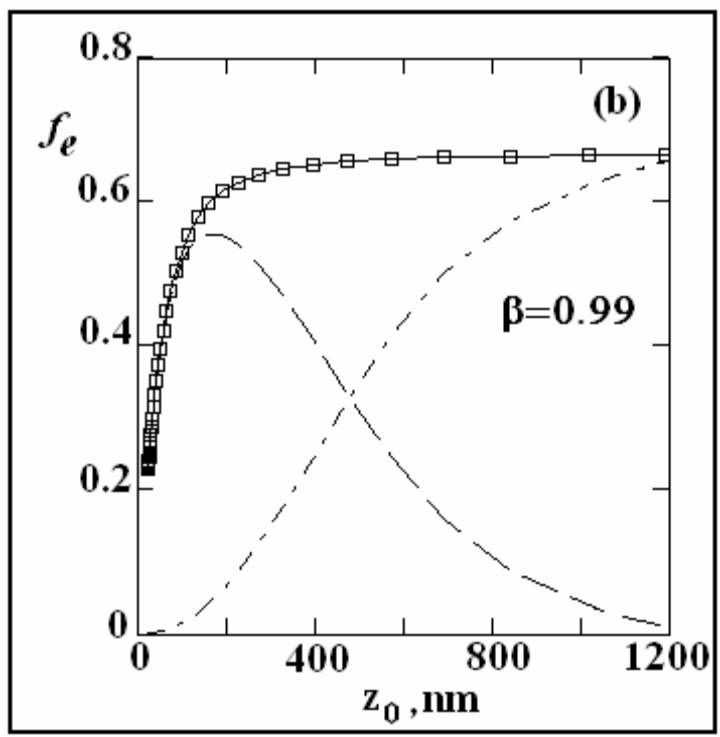

Fig. 1c

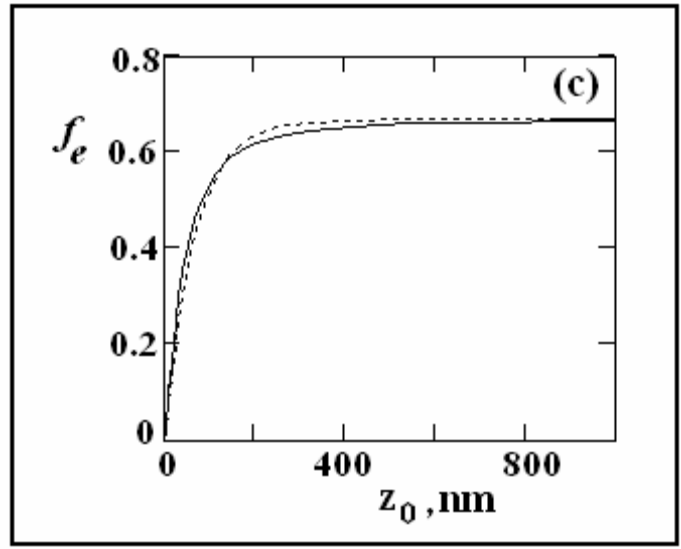

Fig.2a

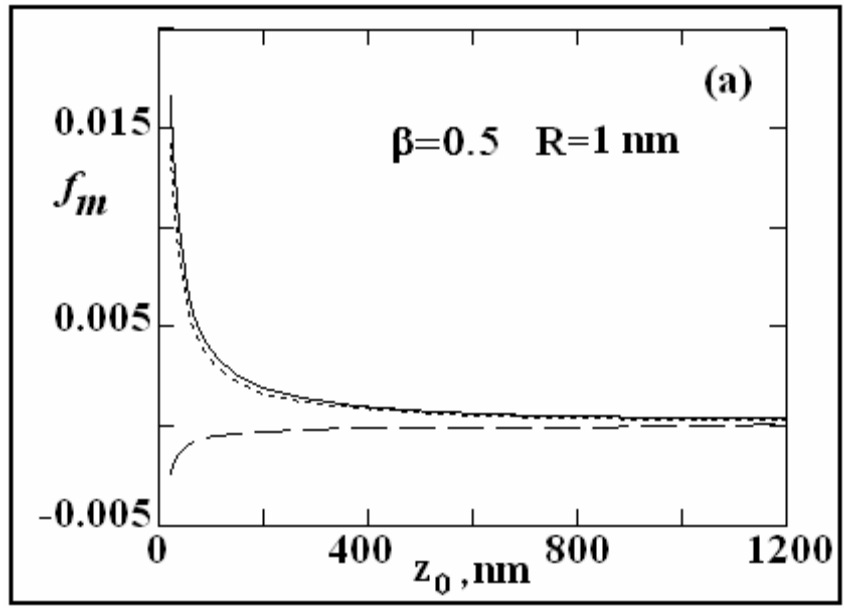

Fig.2b

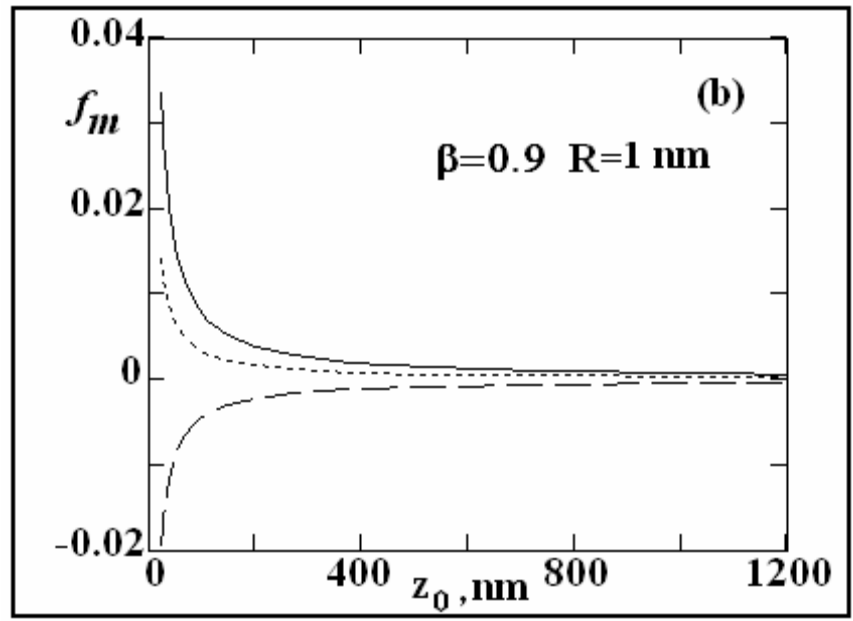

Fig.2c

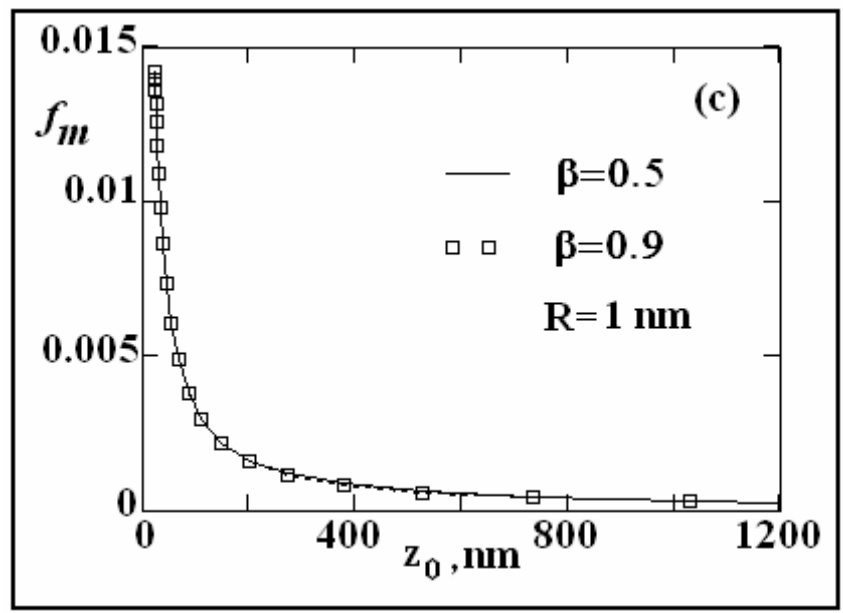

Fig.3a

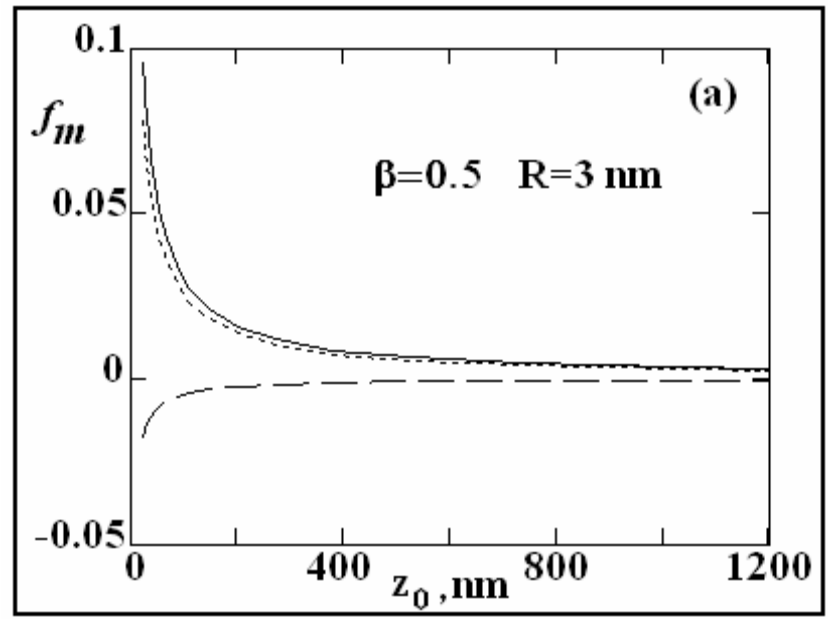

Fig.3b

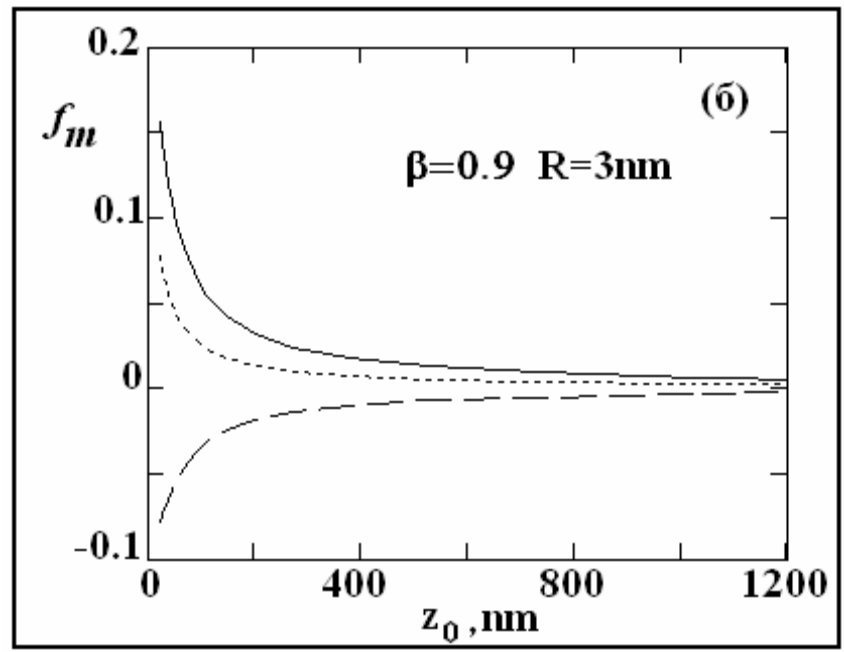

Fig. 3c

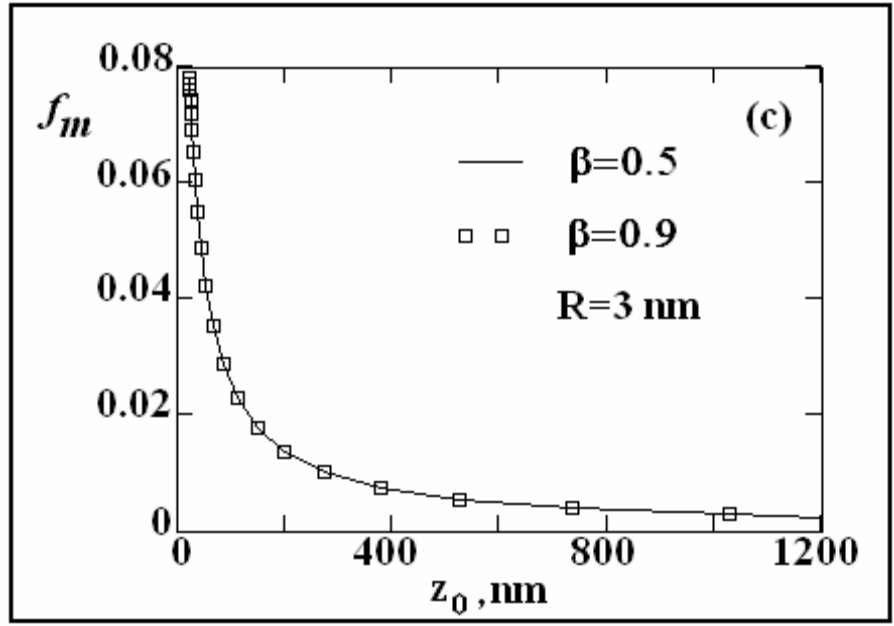

Fig. 4

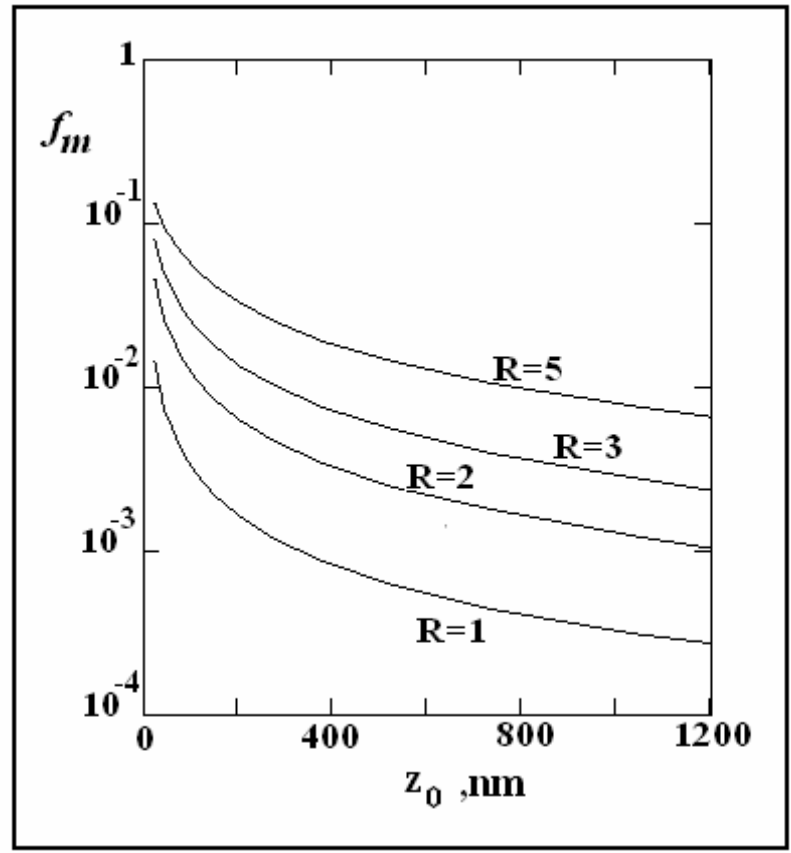